\documentclass[12pt]{article}
\usepackage{amssymb,amsmath,epsfig}

\begin{document}

\title{\bf Dynamics of Potentials in Bianchi Type Scalar-Tensor Cosmology}
\author{M. Sharif \thanks{msharif.math@pu.edu.pk} and Saira Waheed
\thanks{smathematics@hotmail.com}\\
Department of Mathematics, University of the Punjab,\\
Quaid-e-Azam Campus, Lahore-54590, Pakistan.}

\date{}

\maketitle
\begin{abstract}
The present study investigates the nature of the field potential via
new technique known as reconstruction method for the scalar field
potentials. The key point of this technique is the assumption that
Hubble parameter is dependent on the scalar field. We consider
Bianchi type I universe in the gravitational framework of
scalar-tensor gravity and explore the general form of the scalar
field potential. In particular, this field potential is investigated
for the matter contents like barotropic fluid, the cosmological
constant and Chaplygin gas. It is concluded that for a given value
of Hubble parameter, one can reconstruct the scalar potentials which
can generate the cosmology motivated by these matter contents.
\end{abstract}

\section{Introduction}

The reality of cryptic dominant component of the universe
distribution labeled as dark energy (DE) and its resulting phenomena
of cosmic acceleration has become a center of interest for the
researchers. The existence of this unusual sort of DE is supported
by the observational results of many astronomical experiments like
Supernova (Ia) \cite{1N,1}, Wilkinson Microwave Anisotropy Probe
(WMAP) \cite{1*} and Sloan Digital Sky Survey (SDSS) \cite{1**},
galactic cluster emission of X-rays \cite{1+}, large scale-structure
\cite{1++} and weak lensing \cite{1+++}. These experiments reveal
the present day cosmic acceleration by evaluating the luminosity
distance relation of some type of objects known as standard candles.
They also lead to the conclusion that our universe is nearly flat.

In order to resolve these issues, numerous attempts are made which
can be categorized on the basis of the used technique. Basically,
two approaches have been reported in this context: the modification
in the matter configuration of the Lagrangian density and the
modification in the whole gravitational framework described by the
action. The Chaplygin gas \cite{2} and its modified forms \cite{2*},
cosmological constant \cite{2**}, tachyon fields \cite{2***},
quintessence \cite{2****}, viscosity effects \cite{2*****} and
k-essence \cite{2******} etc. are some DE candidates belonging to
the first category. The second approach includes examples of
modified theories like $f(R)$ gravity \cite{3}, Gauss-Bonnet gravity
\cite{3*}, $f(T)$ theory \cite{3**}, $f(R,T)$ gravity \cite{3***}
and scalar-tensor theories \cite{3****}. The study of scalar-tensor
theories in the subject of cosmology has a great worth due to its
vast applications and success \cite{4}.

The complete history of the universe from the early inflationary
epoch to the final era of cosmic expansion can successfully be
discussed by using scalar field as DE candidate \cite{5}. Basically,
the alternating gravitational theories are proposed by the inclusion
of some functions or terms as a possible modification of Einstein
gravity that cannot be derived from the fundamental theory. This
raises a question about the appropriate choice of these functions by
checking their cosmological viability. However, the process of
reconstruction provides a way for having a cosmologically viable
choice of these functions. Such a procedure has been adopted by many
researchers \cite{6}-\cite{15}. The reconstruction procedure is not
a new technique as it has a long history for the reconstruction of
DE models. In order to have a better understanding of this
technique, we may refer the readers to study some interesting
earlier papers \cite{15*}. Basically, this technique enables one to
find the form of the scalar field potential as well as scalar field
for a particular value of the Hubble parameter in terms of scale
factor or cosmic time.

It is worth investigating the nature of scalar field potential in
the context of scalar-tensor theories. Using reconstruction
approach, the nature of the field potential for a minimally coupled
scalar-tensor theory has been discussed \cite{6}. The scalar
potentials for tachyon field \cite{7} as well as for solutions
involving two scalar fields \cite{8} have been reconstructed through
this technique. This is also extended to the modified gravitational
frameworks including non-minimal coupled scalar-tensor theories
\cite{9}, Gauss-Bonnet gravity \cite{10}, $F(T)$ theory \cite{11}
and the non-local gravity model \cite{12}. Kamenshchik et al.
\cite{13} used this technique to reconstruct the scalar field
potential for FRW universe in the induced gravity and discussed it
for some types of matter distribution which can reproduce cosmic
evolution. The same authors \cite{14} used superpotential approach
to reconstruct the field potential for FRW model in a non-minimally
coupled scalar-tensor gravity and explored its nature for different
cases like de Sitter and barotropic solutions describing the cosmic
evolution.

In this paper, we discuss the nature of the field potential using
the reconstruction procedure for locally rotationally symmetric
(LRS) Bianchi type I (BI) universe model. The paper is organized as
follows. In the next section, we provide a general discussion of
this technique and explore the form of scalar field potential.
Section \textbf{3} is devoted to study the field potentials using
the barotropic fluid, the cosmological constant and the Chaplygin
gas as matter contents. In the last section, we discuss and conclude
the results.

\section{General Formulation of the Field Potential}

The scalar-tensor gravity is generally determined by the action
\cite{15}
\begin{eqnarray}\label{3}
S=\int\sqrt{-g}[U(\phi)R-\frac{\omega(\phi)}{2}g^{\mu\nu}\phi_{,\mu}\phi_{,\nu}+V(\phi)]d^4x;
\quad \mu,\nu=0,1,2,3,
\end{eqnarray}
where $U$ is the coupling of geometry and the scalar field, $V$ is
the self-interacting potential, $R$ is the Ricci scalar and $\omega$
is the interaction function. We can discuss different cases of
scalar-tensor theories by taking different values of $U(\phi)$. When
both $U,~\omega$ are constants, the above action yields the
Einstein-Hilbert action with quintessence scalar field, for $U=\phi$
with $\omega=\omega_0,~\omega(\phi)$, it corresponds to simple
Brans-Dicke (BD) and the generalized BD gravity with scalar
potential, respectively. For $U(\phi)=\frac{1}{2}\gamma\phi^2$,
where $\gamma$ is any non-zero constant and constant $\omega$, it
leads to the action of the induced gravity. Anisotropic and
spatially homogeneous extension of flat FRW model, BI universe with
the expansion factors $A$ and $B$ is given by the metric \cite{16}
\begin{eqnarray}\label{1}
ds^2=dt^2-A^2(t)dx^2-B^2(t)(dy^2+dz^2)
\end{eqnarray}
and the respective Ricci scalar is
\begin{eqnarray}\nonumber
R=-2[\frac{\ddot{A}}{A}+2\frac{\ddot{B}}{B}+(\frac{\dot{B}}{B})^2+2\frac{\dot{A}}{A}\frac{\dot{B}}{B}].
\end{eqnarray}
The average scale factor $a(t)$, the universe volume $V$, the
directional Hubble parameters ($H_1$ along $x$ direction while $H_2$
along $y$ and $z$ directions) and the mean Hubble parameter are
given by
\begin{eqnarray}\nonumber
a(t)&=&(AB^{2})^{1/3},\quad V=a^3(t)=AB^{2},\quad
H_{1}=\frac{\dot{A}}{A},\\\nonumber
H_{2}&=&H_{3}=\frac{\dot{B}}{B},\quad
H(t)=\frac{1}{3}(\frac{\dot{A}}{A}+2\frac{\dot{B}}{B}).
\end{eqnarray}

In order to deal with highly non-linear equations, we take a
physical assumption for the scale factors, i.e., $A=B^m;~ m\neq0,1$
\cite{17}. This condition is originated from the fact that in a
spatially homogeneous model, the normal congruence to homogeneous
expansion corresponds to the proportionality of the shear scalar
$\sigma$ and the expansion scalar $\theta$, in other words, the
ratio of these quantities $\frac{\sigma}{\theta}$ is constant. This
condition has been used by many researchers for the discussion of
exact solutions \cite{18}. The above condition further yields the
relations $\frac{\dot{A}}{A}=m\frac{\dot{B}}{B}$ and
$\frac{\ddot{A}}{A}=m\frac{\ddot{B}}{B}+m(m-1)\frac{\dot{B}^2}{B^2}$,
consequently the Ricci scalar takes the form
\begin{eqnarray}\label{2}
R=-2[(m+2)\frac{\ddot{B}}{B}+(m^2+m+1)\frac{\dot{B}^2}{B^2}]
\end{eqnarray}

For BI universe model, we have $\sqrt{-g}=B^{(m+2)}$ and the
respective point-like Lagrangian density constructed by partial
integration \cite{19} of the above action (when $\omega=\omega_0$,
where $\omega_0$ is an arbitrary constant) is given by
\begin{eqnarray}\nonumber
\mathcal{L}(B,\phi,\dot{B},\dot{\phi})&=&2(m+2)B^{(m+1)}\frac{dU}{d\phi}\dot{B}\dot{\phi}
+2B^m\dot{B}^2(1+2m)U(\phi)\\\label{4}
&-&\frac{\omega_0}{2}B^{m+2}\dot{\phi}^2+V(\phi)B^{m+2},
\end{eqnarray}
where we have neglected the boundary terms. In order to formulate
the corresponding field equations, we use the Euler-Lagrange
equations
\begin{eqnarray}\nonumber
\frac{\partial\mathcal{L}}{\partial
B}-\frac{d}{dt}(\frac{\partial\mathcal{L}}{\partial\dot{B}})=0,\quad
\frac{\partial\mathcal{L}}{\partial\phi}-\frac{d}{dt}(\frac{\partial\mathcal{L}}{\partial\dot{\phi}})=0,
\end{eqnarray}
which describe the dependent field equation for the BI model and the
evolution equation of scalar field. Thus we have
\begin{eqnarray}\nonumber
&&2(m+2)\frac{d^2U}{d\phi^2}\dot{\phi}^2-2(m+2)\frac{dU}{d\phi}\ddot{\phi}
-4(1+2m)\frac{dU}{d\phi}\frac{\dot{B}}{B}\dot{\phi}\\\label{6}
&&-4(1+2m)U(\phi)\frac{\ddot{B}}{B}=0,
\end{eqnarray}
\begin{eqnarray}\nonumber
&&\omega_0\ddot{\phi}+\omega_0(m+2)\dot{\phi}\frac{\dot{B}}{B}+2(1+2m)\frac{dU}{d\phi}\frac{\dot{B}^2}{B^2}
-2(m+2)(m+1)\frac{dU}{d\phi}\frac{dV}{d\phi}\frac{\dot{B}^2}{B^2}\\\label{7}
&&-2(m+2)\frac{\ddot{B}}{B}\frac{dU}{d\phi}=0.
\end{eqnarray}
The energy relation (conserved quantity) \cite{20} for the
Lagrangian density (\ref{4}) can be written as
$E_{\mathcal{L}}=\dot{B}\frac{\partial\mathcal{L}}{\partial\dot{B}}
+\dot{\phi}\frac{\partial\mathcal{L}}{\partial\dot{\phi}}-\mathcal{L}$
that yields the independent field equation for BI universe (when
substituted equal to zero)
\begin{eqnarray}\label{8}
2(1+2m)U(\phi)\frac{\dot{B}^2}{B^2}+2(m+2)\frac{dU}{d\phi}\frac{\dot{B}}{B}\dot{\phi}
-\frac{\omega_0}{2}\dot{\phi}^2-V(\phi)=0.
\end{eqnarray}
When $m=1$, these equations reduce to the case of FRW universe
\cite{13}.

For the special choice of $U$, we evaluate the scalar potential in
terms of scale factor, directional Hubble parameter and scalar
field. We consider the directional Hubble parameter as a function of
scale factor or cosmic time by taking different cases of matter
contents. The scalar field is found as a function of scale factor or
cosmic time and then the scale factor as a function of scalar field
by inverting the obtained expression. Finally, we evaluate the
Hubble parameter in terms of scalar field and hence the form of
scalar potential. We shall explore the nature of the potential that
can generate the cosmic evolution described by these matter
contents. Equation (\ref{8}) yields
\begin{eqnarray}\nonumber
V(\phi)=2(1+2m)U(\phi)\frac{\dot{B}^2}{B^2}+2(m+2)\dot{\phi}\frac{dU}{d\phi}
\frac{\dot{B}}{B}-\frac{\omega_0}{2}\dot{\phi}^2
\end{eqnarray}
or equivalently,
\begin{eqnarray}\label{9}
V(\phi)=[2(1+2m)U(\phi)+2(m+2)\phi_{,B}B\frac{dU}{d\phi}-\frac{\omega_0}{2}\phi_{,B}^2B^2]H_2^2,
\end{eqnarray}
which provides
\begin{eqnarray}\nonumber
\frac{dV}{d\phi}&=&2(1+2m)H_2^2\frac{dU}{d\phi}+4(1+2m)U(\phi)\frac{H_2\dot{H}_2}{\dot{\phi}}
+2(m+2)H_2\dot{\phi}\frac{d^2U}{d\phi^2}\\\nonumber
&+&2(m+2)H_2\frac{dU}{d\phi}\frac{\ddot{\phi}}{\dot{\phi}}+2(m+2)\dot{H}_2\frac{dU}{d\phi}-\omega_0\ddot{\phi}.
\end{eqnarray}
Using this equation in Eq.(\ref{7}), it follows that
\begin{eqnarray}\nonumber
&&\omega_0(m+2)\dot{\phi}^2-2(m^2+2)\frac{dU}{d\phi}\dot{\phi}H_2+4(1+2m)U\dot{H_2}
+2(m+2)\dot{\phi}^2\frac{d^2U}{d\phi^2}\\\label{10}
&&+2(m+2)\frac{dU}{d\phi}\ddot{\phi}=0.
\end{eqnarray}

We investigate two cases for the coupling function $U$, i.e., when
$U=U_0$, where $U_0$ is a non-zero constant and $U\equiv U(\phi)$.
In the first case, Eq.(\ref{10}) becomes
\begin{eqnarray}\nonumber
\dot{\phi}^2+(\frac{4(1+2m)U_0}{\omega_0(m+2)})\dot{H}_2=0; \quad
\omega_0\neq0, \quad m\neq-2
\end{eqnarray}
For the scalar field in terms of scale factor $B$, we have
\begin{equation}\label{11}
\phi'^2+[\frac{4(1+2m)U_0}{\omega_0(m+2)}]\frac{H_2'}{H_2B}=0,
\end{equation}
where prime indicates derivative with respect to scale factor,
yielding solution
\begin{equation}\\\nonumber
\phi(B)=\int(\pm\frac{\sqrt{-H_2(B)B(\frac{4(1+2m)U_0}{\omega_0(m+2)})\frac{dH_2}{dB}}}{H_2(B)B})dB+c_1,
\end{equation}
where $c_1$ is a constant of integration. One can solve this
integral for particular values of the Hubble parameter. In the
second case, we consider $U\equiv U(\phi)$ (a non-minimal coupling
of geometry and scalar field). Equation (\ref{10}) can be written
for scalar field in terms of scale factor and directional Hubble
parameter as
\begin{eqnarray}\label{12}
\phi''+\phi'(\frac{H_2'}{H_2})+\phi'^2[\frac{\omega_0/2+\frac{d^2U}{d\phi^2}}{\frac{dU}{d\phi}}]
+\frac{2(1+2m)U}{(m+2)B\frac{dU}{d\phi}}\frac{H_2'}{H_2}+\frac{m(1-m)}{(m+2)}\frac{\phi'}{B}=0.
\end{eqnarray}
This equation is discussed for two particular choices of $U$.

When $U=\phi$, i.e., the simple BD gravity, it follows that
\begin{eqnarray}\label{13}
\phi''+\phi'(\frac{H_2'}{H_2})+\frac{\omega_0}{2}\phi'^2+\frac{2(1+2m)\phi}{(m+2)B}\frac{H_2'}{H_2}
+\frac{m(1-m)}{(m+2)}\frac{\phi'}{B}=0.
\end{eqnarray}
For the case of induced gravity described by
$U(\phi)=\frac{1}{2}\gamma\phi^2$, Eq.(\ref{12}) yields
\begin{eqnarray}\label{14}
\phi''+\phi'(\frac{H_2'}{H_2})+\phi'^2[\frac{\omega_0/2+\gamma}{\gamma\phi}]
+\frac{(1+2m)\gamma\phi}{(m+2)B}\frac{H_2'}{H_2}+\frac{m(1-m)}{(m+2)}\frac{\phi'}{B}=0.
\end{eqnarray}
These two equations are difficult to solve analytically unless the
function $H_2(B)$ is given. For the sake of simplicity, we introduce
a new variable $x\equiv\frac{\phi'}{\phi}$ which yields
$\frac{\phi''}{\phi}=x'+x^2$ and hence Eq.(\ref{14}) turns out to be
\begin{eqnarray}\label{15}
x'+x^2(\frac{2\gamma+\frac{\omega_0}{2}}{\gamma})+x\frac{H_2'}{H_2}
+\frac{(1+2m)}{(m+2)B}\frac{H_2'}{H_2}+\frac{m(1-m)x}{(m+2)B}=0.
\end{eqnarray}
Further, we assume
$x\equiv\frac{2\gamma}{\omega_0+4\gamma}\frac{f'}{f}$, where $f$ is
an arbitrary function of the scale factor $B$. Also,
$x=\frac{\phi'}{\phi}$ thus integration leads to
$\phi=f^{2\gamma/(\omega_0+4\gamma)}$. Using this value of $x$ in
Eq.(\ref{15}), we obtain
\begin{eqnarray}\label{16}
f''+f'\frac{H_2'}{H_2}+\frac{\omega_0+4\gamma}{2\gamma}(\frac{1+2m}{m+2})
\frac{f}{B}\frac{H_2'}{H_2}+\frac{m(1-m)f'}{(m+2)B}=0.
\end{eqnarray}
We see that Eq.(\ref{13}) is difficult to transform in $x$ by the
above transformation. If we consider the scalar field as a constant
then Eq.(\ref{9}) yields the scalar potential $V=2(1+m)U H^2_{2,0}$,
where $H^2_{2,0}$ is constant directional Hubble parameter.
Multiplying the Klein-Gordon equation (\ref{7}) both sides with this
value of $V$, we obtain the scalar potential
$$V=V_0U^{\frac{m^2+2m+3}{1+m}}=V_0\frac{\gamma}{2}\phi^{\frac{2(m^2+2m+3)}{1+m}},$$
which is obviously a constant (as $V_0$ and $\phi$ are constants).

When $\omega\equiv\omega(\phi)$, the field equations (\ref{6}) and
(\ref{8}) remain the same except that the constant $\omega_0$ is
replaced by $\omega(\phi)$ while Eqs.(\ref{7}) becomes
\begin{eqnarray}\nonumber
&&\omega(\phi)\ddot{\phi}+\omega(\phi)(m+2)\dot{\phi}\frac{\dot{B}}{B}
+\frac{\dot{\phi}^2}{2}\frac{d\omega}{d\phi}+2(1+2m)\frac{dU}{d\phi}\frac{\dot{B}^2}{B^2}\\\label{1*}
&&-2(m+2)(m+1)\frac{dU}{d\phi}\frac{dV}{d\phi}\frac{\dot{B}^2}{B^2}-2(m+2)\frac{\ddot{B}}{B}\frac{dU}{d\phi}=0.
\end{eqnarray}
Solving the field equations (\ref{6}), (\ref{8}) and (\ref{1*}), we
have the same expressions as Eqs.(\ref{11}), (\ref{13}) and
(\ref{16}) except $\omega_0$ is replaced by $\omega(\phi)$. In the
following, we discuss Eqs.(\ref{11}), (\ref{13}) and (\ref{16})
separately to construct potential.

\section{Potential Construction}

Now we discuss the scalar field potential by taking three different
matter contents.

\subsection{Barotropic Fluid}

First we consider the barotropic fluid (a particular case of the
perfect fluid) with equation of state (EoS), $p=k\rho,~ 0<k<1,$
where $p$ and $\rho$ are pressure and density, while $k$ is the EoS
parameter. In order to find the evolution of Hubble parameter due to
barotropic fluid, we consider the Einstein field equations for BI
universe model as
\begin{eqnarray}\label{17}
(1+2m)H_2^2=\rho, \quad
(\frac{m+3}{2})\dot{H}_2+(\frac{m^2+m+4}{2})H_2^2=-p,
\end{eqnarray}
where we have used the condition $A=B^m$ and also combined the two
dependent field equations. The integration of the energy
conservation equation yields $\rho=\rho_0B^{-(1+k)(m+2)}$, where
$\rho_0$ is an integration constant. Consequently, the directional
Hubble parameters are found to be
\begin{eqnarray}\nonumber
H_2(B)&=&\frac{H_1(B)}{m}=[(\frac{m^2+m+4}{1+2m}+2k)
\frac{2\rho_0}{(1+k)(m+2)(m+3)}]^{1/2}\\\label{18}
&\times&B^{-\frac{(1+k)(m+2)}{2}},
\end{eqnarray}
where the integration constant is taken to be zero. The evolution of
Hubble parameter is $\frac{H'_2(B)}{H_2(B)}=-\frac{(1+k)(m+2)}{2B}$.
The corresponding deceleration parameter turns out to be positive,
i.e., $q=-1+\frac{3(k+1)}{2}$ which is consistent with the
barotropic fluid. Using these values in Eq.(\ref{11}), we obtain
$$\phi(B)=\ln(\phi_0B^{\pm\sqrt{\frac{2(1+k)(1+2m)U_0}{\omega_0}}}),$$
where $\phi_0$ is a non-zero integration constant. This shows that
the constant coupling of geometry and scalar field, i.e., $U=U_0$
for the barotropic fluid leads to the logarithmic form of scalar
field which further corresponds to expanding or contracting scalar
field versus scale factor $B$ on the basis of sign. Consequently,
the scale factors turn out to be
$$A(\phi)=(\frac{\exp(\phi)}{\phi_0})^{\mp
m\sqrt{\frac{2(1+k)(1+2m)U_0}{\omega_0}}},\quad
B(\phi)=(\frac{\exp(\phi)}{\phi_0})
^{\mp\sqrt{\frac{2(1+k)(1+2m)U_0}{\omega_0}}}.$$ We see that the
scale factors are of exponential form which indicate rapid cosmic
expansion for the expanding scalar field. The corresponding field
potential is
\begin{eqnarray}\nonumber
V(B)&=&[2(1+2m)U_0-(1+k)(1+2m)U_0](\frac{m^2+m+4}{1+2m}+2k)\\\label{18*}
&\times&\frac{2\rho_0}{(1+k)(m+2)(m+3)}B^{-(1+k)(m+2)}.
\end{eqnarray}
This is of power law nature and indicates inverse power law behavior
for $m>0$ as $0<k<1$.

For the variable $\omega(\phi)$, we consider the ansatz
$\omega(\phi)=\omega_0\phi^n;~n>0$ so that the scalar field takes
the following form
\begin{eqnarray}\nonumber
\phi(B)&=&[c_2(\ln(B)^2n^2+4\ln(B)^2n-2\ln(B)n^2c_1-8\ln(B)nc_1+4\ln(B)^2\\\nonumber
&-&8\ln(B)c_1+c_1^2n^2+4nc_1^2+4c_1^2)]^{1/(n+2)}(2^{\frac{1}{n+2}})^{-2},
\end{eqnarray}
where $c_1$ is an integration constant and
$c_2=\frac{2(1+2m)(1+k)U_0}{\omega_0}$. For the sake of simplicity,
we take $c_1=0$ and hence the scalar field becomes
$$\phi(B)=\frac{c_2^{1/(n+2)}(\ln(B^{2n^2+8n+8}))^{1/(n+2)}}{2^{(1/(n+2))^2}}.$$
Thus the scale factors in exponential form are
$$A(\phi)=\exp(\frac{4m}{(2n^2+8n+8)}c_2^{-1}\phi^{n+2}),\quad
B(\phi)=\exp(\frac{4}{2n^2+8n+8}c_2^{-1}\phi^{n+2}).$$ Consequently,
the potential turns out to be
\begin{eqnarray}\nonumber
V(B)&=&[2(1+2m)U_0-\frac{\omega_0}{2}(\frac{c_2^{1/(n+2)}\ln(B^{2n^2+8n+n})}{2^{(1/(n+2))^2}})^nc_2^{2/(n+2)}
(2^{1/(n+2)})^{-4}\\\nonumber
&\times&\frac{(2n^2+8n+8)^2}{(n+2)^2}(\ln(B^{2n^2+8n+n}))^{-2\frac{(1+n)}
{n+2}}](\frac{m^2+m+4}{1+2m}+2k)\\\label{18**}
&\times&\frac{2\rho_0}{(1+k)(m+2)(m+3)}B^{-(1+k)(m+2)},
\end{eqnarray}
which contains the product of inverse power law and logarithmic
functions of the scale factor.

For $U=\phi$, Eq.(\ref{13}) takes the form
\begin{eqnarray}\nonumber
\phi''+(\frac{m(1-m)}{m+2}-\frac{(1+k)(m+2)}{2})\frac{\phi'}{B}
+\frac{\omega}{2}\phi'^2-(1+2m)(1+k)\frac{\phi}{B^2}=0.
\end{eqnarray}
When $\omega=\omega_0$ or $\omega(\phi)=\omega_0\phi^n$, the
solution to this differential equation is quite complicated and
cannot provide much insights. However, if we take $m=-1/2$ and
$\omega=\omega_0$, then this leads to
\begin{eqnarray}\label{*}
\phi(B)=\frac{2}{\omega_0}\ln[\frac{\frac{\omega_0}{6}(4c_3B^{3/4(3+k)}+9c_4+3c_4k)}{3+k}],
\end{eqnarray}
where $c_3$ and $c_4$ are integration constants and $\omega_0\neq0$.
The respective scale factors are
\begin{eqnarray*}\nonumber
A(\phi)&=&[\frac{1}{4c_3}(\frac{6}{\omega_0}\exp(\frac{\omega_0}{2}\phi)-9c_4-3c_4k)]^{m/(3/4(3+k))},\\
B(\phi)&=&[\frac{1}{4c_3}(\frac{6}{\omega_0}\exp(\frac{\omega_0}{2}\phi)-9c_4-3c_4k)]^{1/(3/4(3+k))}
\end{eqnarray*}
and the corresponding scalar field potential turns out to be
\begin{eqnarray}\nonumber
V(B)&=&2(m+2)(\frac{3c_3(3+k)B^{3/4(3+k)}}{\frac{\omega_0}{2}
(4c_3B^{3/4(3+k)}+9c_4+3c_4k)})\\\label{a}
&-&\frac{\omega_0}{2}(\frac{3c_3(3+k)B^{3/4(3+k)}}{\frac{\omega_0}{2}
(4c_3B^{3/4(3+k)}+9c_4+3c_4k)})^2.
\end{eqnarray}
We can conclude that the scalar field is described by logarithmic
function and the scale factors are of exponential nature which
yields expansion for increasing scalar field while the potential
turns out to be of power law nature.

Now we discuss the induced gravity case and evaluate the function
$f$ by using the Hubble parameter and its evolution in Eq.(\ref{16})
which leads to
\begin{eqnarray}\nonumber
f''+[\frac{m(1-m)}{m+2}-\frac{(1+k)(m+2)}{2}]\frac{f'}{B}
-\frac{\omega_0+4\gamma}{4\gamma}(1+2m)(1+k)\frac{f}{B^2}=0
\end{eqnarray}
whose solution is
\begin{eqnarray}
f(B)=c_5B^{r_1}+c_6B^{r_2}; \quad
r_{1,2}=\frac{1-c_7}{2}\pm\frac{1}{2}\sqrt{c_7^2+1-2c_7-4c_8},
\end{eqnarray}
where $c_5$ and $c_6$ are arbitrary constants while $c_7$ and $c_8$
are given by
\begin{equation}\nonumber
c_7=-\frac{2m+(3+k)m^2+4+4(1+m)k}{2(m+2)},~
c_8=-(1+2m)(1+k)\frac{\omega_0+4\gamma}{4\gamma}.\label{19}
\end{equation}
The corresponding scalar field is
$\phi(B)=(c_5B^{r_1}+c_6B^{r_2})^{\frac{2\gamma}{\omega_0+4\gamma}}$
which is clearly of power law nature. Since it is difficult to
invert this expression for the scale factor $B$ in terms of $\phi$,
so we take either $c_5=0$ or $c_6=0$, which leads to either
$$A(\phi)=\frac{1}{c_5^m}\phi^{\frac{m(\omega_0+4\gamma)}{4r_1\gamma}},
\quad
B(\phi)=\frac{1}{c_5}\phi^{\frac{\omega_0+4\gamma}{4r_1\gamma}},$$
or
$$A(\phi)=\frac{1}{c_6^m}\phi^{\frac{m(\omega_0+4\gamma)}{4r_2\gamma}},
\quad
B(\phi)=\frac{1}{c_6}\phi^{\frac{\omega_0+4\gamma}{4r_2\gamma}}.$$
We see that the scale factors are also of power law nature and show
expanding or contracting behavior depending upon the values of the
involved parameters. The scalar field potential (\ref{9}) then turns
out to be
\begin{eqnarray}\nonumber
V(B)&=&[(1+2m)\gamma+\frac{4\gamma^2(m+2)r_{1,2}}{\omega_0+4\gamma}-\frac{2\omega_0
\gamma^2r_{1,2}^2}{(\omega_0+4\gamma)^2}](\frac{m^2+m+4}{1+2m}+2k)\\\label{20}
&\times&(\frac{\rho_0c_{5,6}^{\frac{4\gamma}{\omega_0+4\gamma}}}{(1+k)(m+2)})B^{\frac{4\gamma
r_{1,2}}{\omega_0+4\gamma}-(1+k)(m+2)}.
\end{eqnarray}
This may be of positive or inverse power law nature depending upon
the values of parameters.
\begin{figure}
\centering \epsfig{file=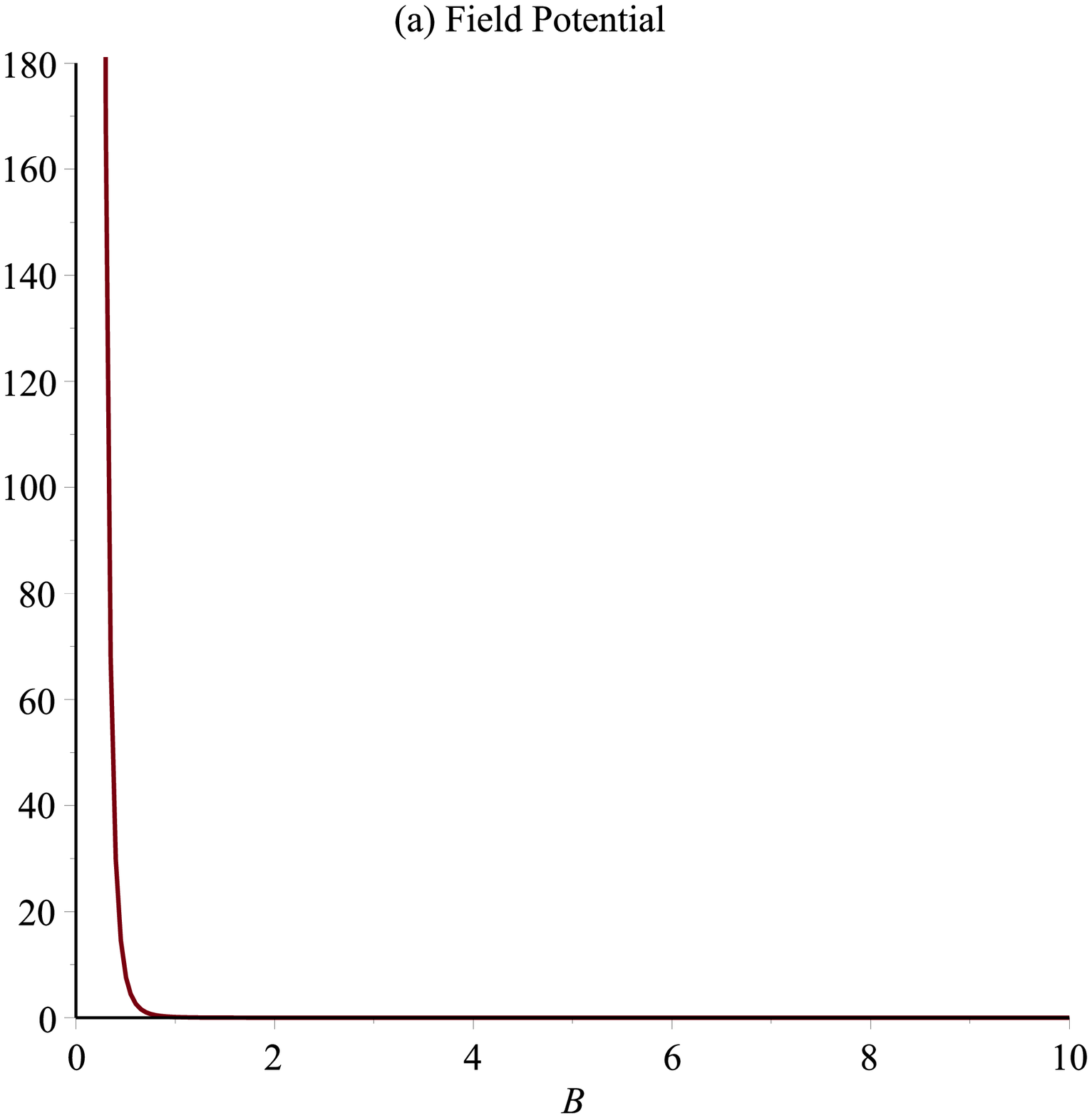,width=.45\linewidth}
\epsfig{file=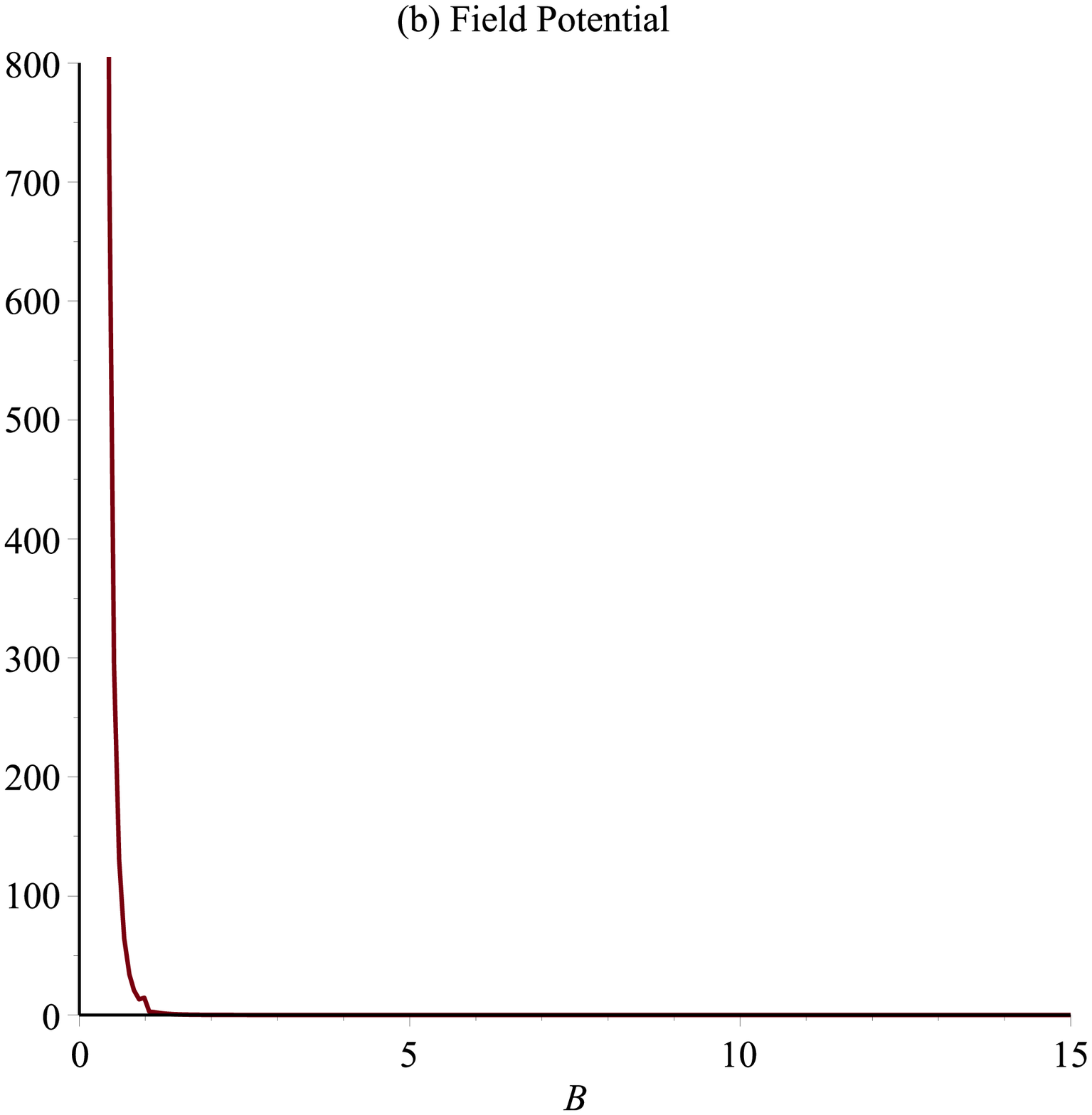,width=.45\linewidth}\\
\centering \epsfig{file=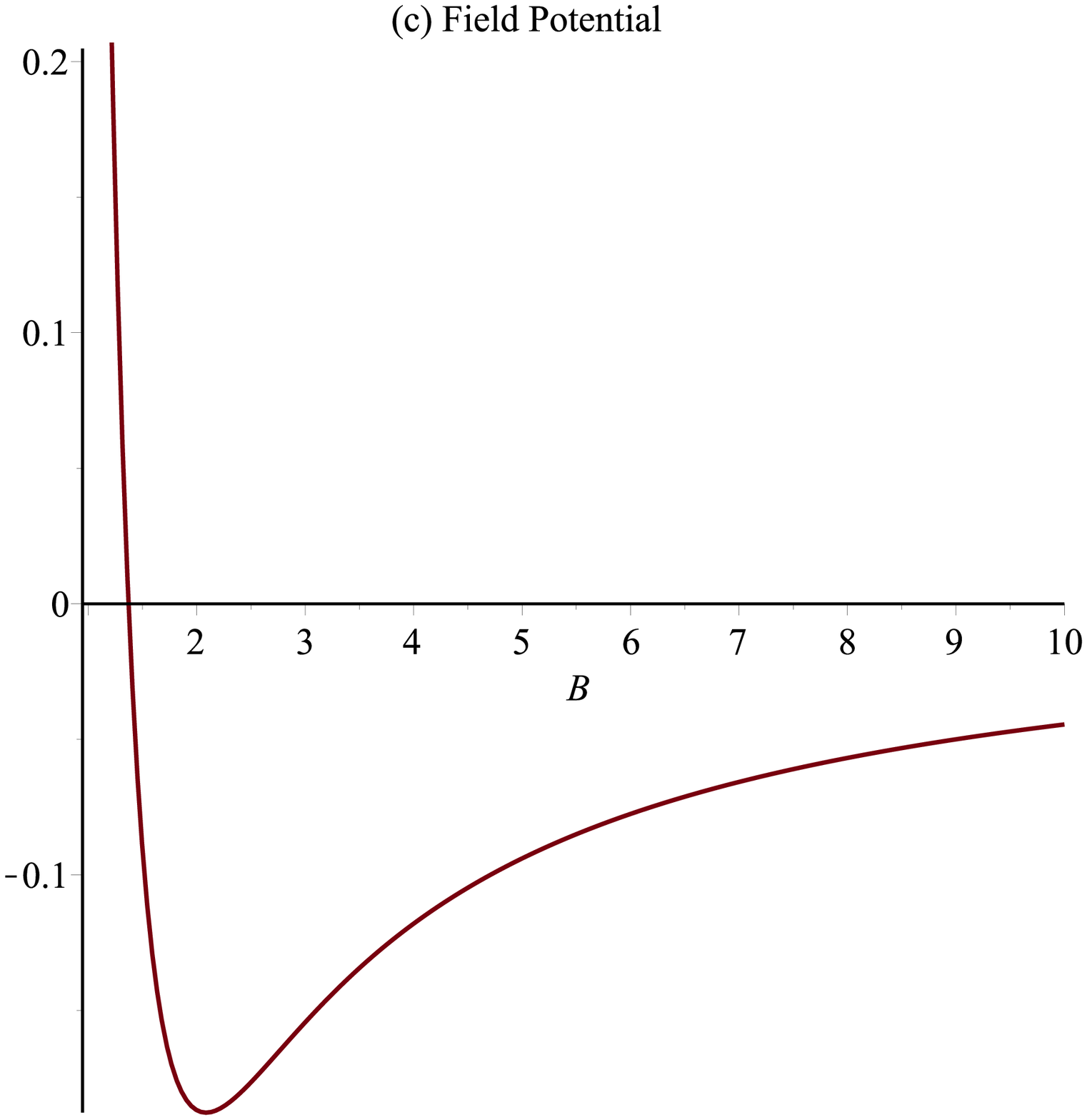,width=.45\linewidth}
\epsfig{file=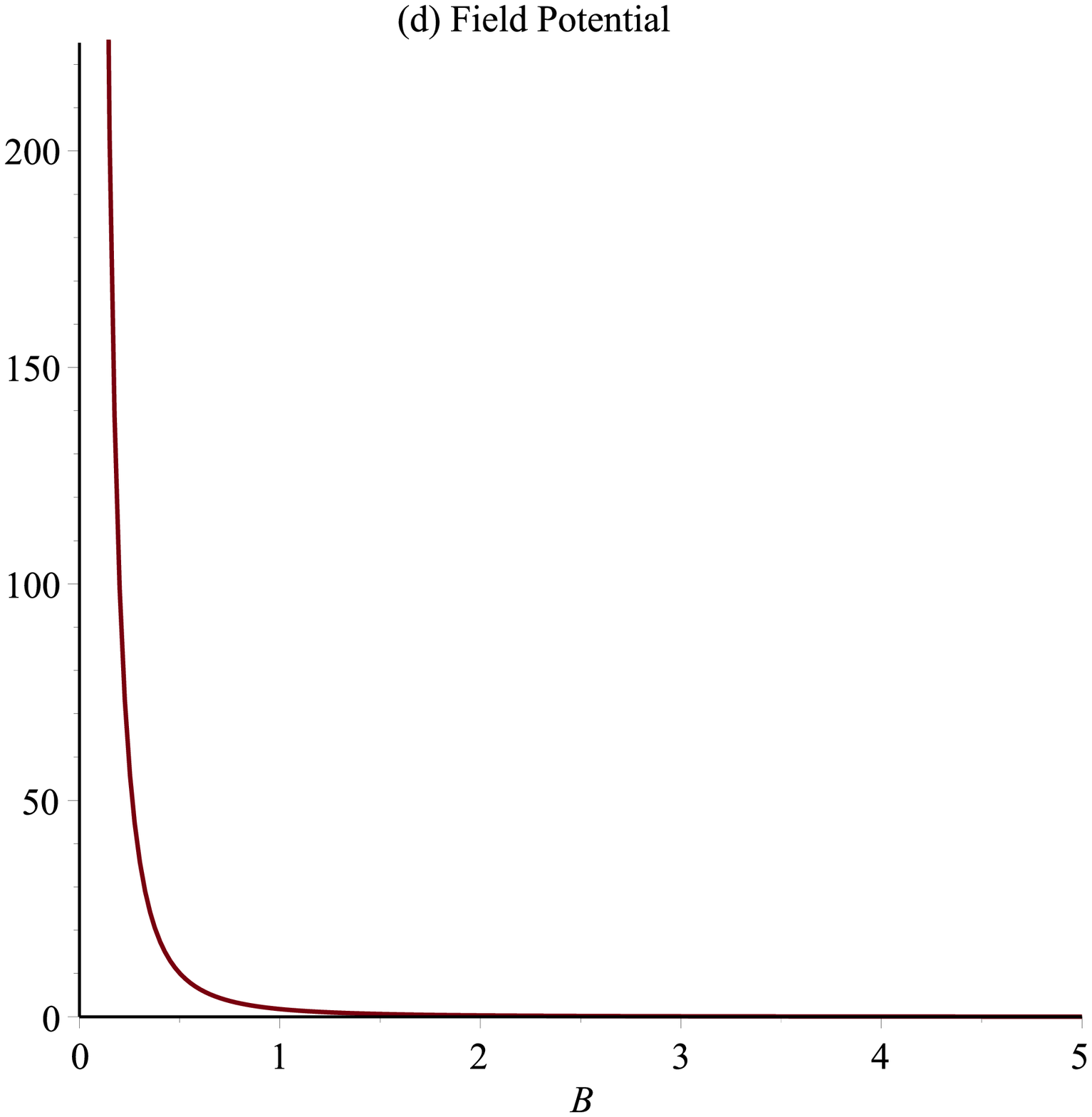,width=.45\linewidth} \caption{Plots show the
field potential versus scale factor $B$. Plots (a), (b), (c) and (d)
correspond to the field potentials given by Eqs.(\ref{18*}),
(\ref{18**}), (\ref{a}) and (\ref{20}), respectively. Here
$m=2,~\rho_0=1,~U_0=3,~k=0.5$ and $\omega_0=0.9$ in all plots except
for the plot (c), where $m=-0.5$.}
\end{figure}

For variable $\omega$, the analytical solution of Eq.(\ref{16}) is
not possible. However, the corresponding numerical solution can be
found by using the initial conditions $f(1)=0.67$ and $f'(1)=1.95$
and is given by the polynomial interpolation
\begin{eqnarray}\nonumber
&&f(B)=0.014B^8-0.4615B^7+6.3599B^6-47.5667B^5+206.9123B^4\\\label{b}
&&-529.2141B^3+772.0721B^2-587.8872B+180.4408,
\end{eqnarray}
where we have taken $m=2,~\gamma=0.25,~k=0.5$ and
$\omega=0.9\phi^2$. The corresponding scalar field is
$\phi(B)=(f(B))^{\frac{2\gamma}{\omega_0+4\gamma}}$,  yielding the
form of the field potential in polynomial form which represents
positive power law nature. Here the scalar field is in polynomial
form which cannot be inverted for scale factor $B$.

We have plotted the potentials given by Eqs.(\ref{18*}),
(\ref{18**}), (\ref{a}) and (\ref{20}) versus scale factor $B$ as
shown in Figure \textbf{1}. It is found that in all cases, the
scalar field potentials are positive decreasing functions except for
the plot (c) which has a signature flip from positive to negative
with the increase in scale factor (this graph corresponds to the
negative value of $m$). We can conclude that for a positive behavior
of the field potential (which is physically acceptable), we should
take positive range of $m$.

\subsection{Cosmological Constant}

In this case, we take $p=-\rho$ and hence the energy density becomes
a constant, i.e., $\rho=\rho_0$. The corresponding directional
Hubble parameters and its evolution are given by
\begin{eqnarray}\label{21}
\frac{H_1(B)}{m}=H_2(B)=\sqrt{\frac{4\rho_0}{m+3}(1-\frac{m^2+m+4}{1+2m})B},
\quad \frac{H'_2(B)}{H_2(B)}=\frac{1}{2B\ln(B)}.
\end{eqnarray}
The deceleration parameter turns out to be a dynamical quantity
$q=-(1+\frac{1}{2(m+2)\ln(B)})$. It is interesting to mention here
that in our case, the directional Hubble parameters are dependent on
the scale factor $B$ (due to anisotropy) whereas in the case of FRW
universe, the Hubble parameter is independent of the scale factor,
i.e., it turns out to be constant. We use these values in the
previously discussed three cases, i.e., $U=U_0,~\phi$ and
$U=\frac{1}{2}\gamma\phi^2$. Equation (\ref{11}) provides
$(\phi')^2=\frac{(1+2m)U_0}{\omega_0(m+2)}\frac{1}{B^2\ln(B)}$ whose
integration leads to $\phi(B)=\pm\sqrt{-2\ln(B)c_{10}}+c_9$, where
$c_9$ is an integration constant while
$c_{10}=\frac{2(1+2m)U_0}{\omega_0(m+2)}$. This leads to the scale
factor as an exponential function of the scalar field
$B(\phi)=\exp(-1/2c_{10}(\phi-c_9))$. Likewise, for
$\omega=\omega_0\phi^n$, the scalar field is found to be
\begin{eqnarray}\label{2*}
\phi(B)=(2^{-2/(n-2)})^2[\frac{\pm\sqrt{-2\ln(B)c_{10}}+c_{11}}{(n-2)(2\ln(B)c_{10}+c_{11}^2)}],
\end{eqnarray}
where $c_{11}$ is an integration constant while $c_{10}$ is the same
as above. Using these values in Eq.(\ref{9}), the field potential
can be determined which would include the product terms of scale
factor and logarithmic function.

In the case of simple BD gravity, Eq.(\ref{13}) is not easy to solve
for both cases $\omega=\omega_0$ and $\omega=\omega_0\phi^n$.
However, the corresponding numerical solutions can be constructed in
a similar way as we have discussed in the previous case. The scalar
field as well as the potentials constructed, in this way, would be
of polynomial nature. For $m=-1/2$, it leads to
$\phi''+\frac{\omega_0}{2}\phi'=0$ and hence
$$\phi(B)=\frac{2\ln(\frac{c_{12}}{2}B\omega_0+\frac{1}{2}c_{13}\omega_0)}{\omega_0},\quad
B(\phi)=\frac{2}{c_{12}\omega_0}(\exp(\omega_0\phi/2)-\frac{c_{13}\omega_0}{2}),$$
where $c_{12}$ and $c_{13}$ are integration constants. The field
potential corresponding to these values can be obtained from
Eq.(\ref{9}) which would be of power law nature. For the case of
induced gravity, Eq.(\ref{16}) provides
\begin{eqnarray}
f''+\frac{f'}{2B\ln
B}+\frac{m(1-m)}{m+2}\frac{f'}{B}+\frac{(\omega_0+4\gamma)}{4\gamma}
\frac{(1+2m)}{(m+2)}\frac{f}{B}(\frac{1}{2B\ln(B)})=0.
\end{eqnarray}
Solving this equation, we have the solution in terms of Kummer
functions
\begin{eqnarray}\nonumber
f(B)&=&c_{14}KummerM(\frac{1}{4}(-m(1-m)\gamma+(2\omega_0+9\gamma)m+6\gamma+\omega_0(m+2)\\\nonumber
&+&3(m+2)\gamma(m+2-m(1-m)))((m+2)\gamma(m+2-m(1-m)))^{-1},3/2,\\\nonumber
&&(-m-2+m(1-m))(m+2)\ln(B)
\sqrt{\ln(B)}B^{1/2\frac{(2(m+2)(m+2-m(1-m)))}{(m+2)^2}})\\\nonumber
&+&c_{15}KummerU(\frac{1}{4}(-m(1-m)\gamma+(2\omega_0+9\gamma)m+6\gamma+\omega_0(m+2)\\\nonumber
&+&3(m+2)\gamma(m+2-m(1-m)))((m+2)\gamma(m+2-m(1-m)))^{-1},3/2,\\\nonumber
&&(-m-2+m(1-m))(m+2)\ln(B)\sqrt{\ln(B)}B^{1/2\frac{(2(m+2)(m+2-m(1-m)))}{(m+2)^2}}),\\\label{6*}
\end{eqnarray}
where $c_{14}$ and $c_{15}$ are integration constants. Since
$\phi=f^{2\gamma/(\omega_0+4\gamma)}$, consequently the scalar field
potential can be determined (it would be a lengthy expression in
Kummer function). For $\omega_0=-4\gamma$, the solution is
\begin{equation}\label{5*}
f(B)=c_{16}+(\int \frac{B^{-m(1-m)/m+2}}{\sqrt{\ln(B)}}dB)c_{17},
\end{equation}
where $c_{16}$ and $c_{17}$ are integration constants. The
corresponding potential can be determined by using the value of the
scalar field $\phi=f^{\frac{2\gamma}{\omega_0+4\gamma}}$ in
Eq.(\ref{9}). It would include the integral term and hence cannot be
categorized as power law, exponential or logarithmic form.

\subsection{Chaplygin Gas}

Finally, we consider the Chaplygin gas EoS as DE candidate which is
defined by $p=-\frac{C}{\rho}$, where $C$ is some positive constant.
In order to discuss the potential, we use the above EoS parameter in
the energy conservation equation and then integration leads to
$\rho(B)=(C+c_{18}B^{-2(m+2)})^{1/2}$, where $c_{18}$ is an
integration constant. Using this value in Eq.(\ref{17}), it follows
that
\begin{eqnarray}\nonumber
H_2^2(B)&=&\frac{4C^{1/2}}{m+3}(1-\frac{m^2+m+4}{1+2m})\ln(B)
+\frac{c_{18}}{(m+2)(m+3)C^{1/2}}\\\label{40}&\times&(1+\frac{m^2+m+4}{2(1+2m)})B^{-2(m+2)},
\end{eqnarray}
whose evolution yields
\begin{eqnarray}
\frac{H'_2}{H_2}=\frac{p_1-2(m+2)p_2B^{-2(m+2)}}{2B(p_1\ln(B)+p_2B^{-2(m+2)})},
\end{eqnarray}
where $p_1=\frac{4C^{1/2}}{m+3}(1-\frac{m^2+m+4}{1+2m})$ and
$p_2=\frac{c_{18}}{(m+2)(m+3)\sqrt{C}}(1+\frac{m^2+m+4}{1+2m})$. For
the constant coupling of scalar field and geometry ($U=U_0$) with
$\omega=\omega_0$, we have
\begin{eqnarray}\nonumber
\phi(B)&=&\int
\pm\sqrt{2}(\omega_0(m+2)(B^{-2(1+m)}p_2+B^2p_1\ln(B))U_0(1+2m)(-p_1\\\nonumber
&+&2mp_2B^{-2(m+2)}+4p_2B^{-2(m+2)})((B^{-2(m+2)}p_2+B^2p_1\ln(B))\\\nonumber
&\times&\omega_0(m+2))^{-1})^{1/2}.
\end{eqnarray}
Thus we can determine the field potential that can generate the
cosmic evolution of Chaplygin gas matter (it would be in integral
form). For $\omega=\omega_0\phi^n$, the scalar field is
\begin{eqnarray}\nonumber
&&\frac{2\phi(B)^{(n+2)/2}}{n+2}+\int[(\omega_0(m+2)(p_2+\ln(B')B'^{2m+4}p_1))^{-1}(\phi(B)^{n/2}B'^{2(1+m)}\\\nonumber
&&\times(-2U_0\omega_0(2m^2+5m+2)\phi(B)^{-n}(-2B'-4mp_2^2m-2p_1\ln(B')mp_2B'^{4-2m}\\\nonumber
&&+B'^8p_1^2\ln(B')-4p_1\ln(B')p_2B'^{4-2m}+B'^{4-2m}p_1p_2-4B'^{-4m}p_2^2))^{1/2})B'^{-6}]=0.
\end{eqnarray}
Clearly, it is not possible to have an explicit expression for
scalar field in terms of scale factor $B$ and hence the form of the
respective field potential cannot be determined. For simple BD
gravity with $\omega=\omega_0$ and $\omega=\omega_0\phi^n$, we could
not find analytical solutions but numerical solutions can be
constructed in a similar pattern as we have discussed earlier. For
induced gravity, analytical solution is only possible if we take
$p_2=0$, which further implies the same cases as we have found in
the cosmological constant case (as
$\frac{H'_2}{H_2}=\frac{1}{2B\ln(B)}$).

\section{Summary and Discussion}

This paper investigates scalar field potentials by a new technique
known as the reconstruction technique for the field potentials. We
have applied this technique to BI universe model in the context of
general scalar-tensor theory. The general form of the field
potential without assigning any values of $U,~V$ and $H_2$ has been
explored. We have also discussed two particular cases of $U$, i.e.,
when it is a constant and $U=U(\phi)$. In both cases, the field
potential depends upon the scale factor $B$, the scalar field and
the directional Hubble parameter $H_2$. Further, we have taken two
cases for $\omega$, i.e., $\omega=\omega_0$ and
$\omega=\omega_0\phi^n$. It is found that an explicit form of the
field potential cannot be found in terms of scale factor unless we
choose some particular value of the Hubble parameter. For this
purpose, we have taken the evolution of Hubble parameter motivated
by the barotropic fluid, the cosmological constant and the Chaplygin
gas matter contents. In literature \cite{16, 21}, four types of
scalar field potentials have usually been discussed, i.e., the
positive and inverse power laws, the exponential and the logarithmic
potentials while other forms are multiple of these four types.

For the barotropic fluid, the potential can be found but it is not
possible for the simple BD gravity. We have also observed that for
constant $U$, the scalar fields are logarithmic functions for both
$\omega=\omega_0$ and $\omega=\omega_0\phi^n$, while the scale
factors are of exponential nature. Also, for simple BD gravity with
$m=-0.5$ and $\omega=\omega_0$, the scale factors are exponential
functions while for the induced gravity, they turn out to be of
power law form. In order to examine their behavior, we have plotted
the field potentials versus scale factor $B$ as shown in Figure
\textbf{1}. It is concluded that the field potentials are positive
and decrease to zero except for the case of simple BD gravity where
we have taken negative value of $m$. We may conclude that for
positive field potential, we should impose the condition $m>0$. We
have also discussed a numerical approach (polynomial interpolation)
for the cases where no analytical solution exists. Likewise, for the
cosmological constant candidate of DE with constant coupling
function $U$, we can determine the form of the field potential
without taking any condition for both $\omega$, however in other
cases, we have to impose some certain conditions.

In the case of Chaplygin gas matter contents, the scalar field
potential can be discussed only for $\omega=\omega_0$ with $U=U_0$.
However, in other cases, either the explicit analytical solution is
not possible or we have the same expression of the field potential
as in the case of cosmological constant. It would be worthwhile to
investigate the form of the field potential for the exponential form
of coupling function of scalar field and geometry. This procedure
may lead to some interesting results when the chameleon mechanism is
taken into account in the framework of scalar-tensor gravity.


\vspace{0.25cm}

\begin{thebibliography}{44}

\bibitem{1N} Riess, A.G. et al.: Astrophys. J. \textbf{116}(1998)1009.

\bibitem{1} Perlmutter, S. et al.: Nature \textbf{391}(1998)51.

\bibitem{1*} Bennett, C.L. et al.: Astrophys. J. Suppl. \textbf{148}(2003)1.

\bibitem{1**} Tegmark, M. et al.: Phys. Rev. D \textbf{69}(2004)03501.

\bibitem{1+} Allen, S.W., Schmidt, R.W., Ebeling, H., Fabian, A.C.
and Speybroeck, L.V.: Mon. Not. Roy. Astron. Soc.
\textbf{353}(2004)457.

\bibitem{1++} Hawkins, E. et al.: Mon. Not. Roy. Astr. Soc. \textbf{346}(2003)78.

\bibitem{1+++} Jain, B. and Taylor, A.: Phys. Rev. Lett. \textbf{91}(2003)141302.

\bibitem{2} Chaplygin, S.: Sci. Mem. Moscow Univ. Math. Phys.
\textbf{21}(1904)1.

\bibitem{2*} Bento, M.C., Bertolami, O. and Sen, A.A.: Phys. Rev. D \textbf{66}(2002)043507.

\bibitem{2**} Padmanabhan, T.: Gen. Relativ. Gravit. \textbf{40}(2008)529.

\bibitem{2***} Gorini, V., Kamenshchik, A.Y., Moschella, U. and Pasquier, V.: Phys. Rev. D
\textbf{69}(2004)123512.

\bibitem{2****} Ratra, B. and Peebles, P.J.E.: Phys. Rev. D \textbf{37}(1988)3406.

\bibitem{2*****} Murphy, G.L.: Phys. Rev. D \textbf{8}(1973)4231;
Calvao, M.O., de Oliveira, H.P., Pavon, D. and Salim, J.M.: Phys.
Rev. D \textbf{45}(1992)3869.

\bibitem{2******} Chiba, T., Okabe, T. and Yamaguchi, M.: Phys. Rev. D \textbf{62}(2000)023511.

\bibitem{3} Sotiriou, T.P. and Faraoni, V.: Rev. Mod. Phys. \textbf{82}(2010)451;
Felice, A.D. and Tsujikawa, S.: Living Rev. Rel. \textbf{13}(2010)3.

\bibitem{3*} Elizalde, E., Nojiri, S. and Odintsov,
S.D.: Phys. Rev. D \textbf{70}(2004)043539; Cognola, G., Elizalde,
E., Nojiri, S., Odintsov, S.D. and Zerbini, S.: J. Cosmol.
Astropart. Phys. \textbf{0502}(2005)010; Cognola, G., Elizalde, E.,
Nojiri, S., Odintsov, S.D. and Zerbini, S.: Phys. Rev. D
\textbf{73}(2006)084007; Elizalde, E., Nojiri, S., Odintsov, S.D.,
Sebastiani, L. and Zerbini, S.: Phys. Rev. D
\textbf{83}(2011)086006.

\bibitem{3**} Linder, E.V.: Phys. Rev. D \textbf{81}(2010)127301;
Daouda, M.H., Rodrigues, M.E. and Houndjo, M.J.S.: Eur. Phys. J. C
\textbf{72}(2012)1890.

\bibitem{3***} Harko, T. et al.: Phys. Rev. D \textbf{84}(2011)024020.

\bibitem{3****} Brans, C.H. and Dicke, R.H.: Phys. Rev.
\textbf{124}(1961)925; Faraoni, V.: \emph{Cosmology in Scalar-Tensor
Gravity} (Springer, 2004).

\bibitem{4} Bertolami, O. and Martins, P.J.: Phys. Rev. D \textbf{61}(2000)064007;
Banerjee, N. and Pavon, D.: Phys. Rev. D \textbf{63}(2001)043504.

\bibitem{5} Peterson, C.M. and Tegmark, M.: Phys. Rev. D \textbf{83}(2011)023522;
Pi, S. and Sasaki, M.: J. Cosmol. Astropart. Phys.
\textbf{1210}(2012)051.

\bibitem{6} Capozziello, S., Nojiri, S. and Odintsov, S.D.: Phys. Lett. B
\textbf{634}(2006)93; Kamenshchik, A.Y. and Manti, S.: Gen. Relativ.
Gravit. \textbf{44}(2012)2205.

\bibitem{7} Feinstein, A.: Phys. Rev. D \textbf{66}(2002)063511;
Shchigolev, V.K. and Rotova, M.P.: Phys. Lett. A
\textbf{27}(2012)1250086.

\bibitem{8} Andrianov, A.A., Cannata, F., Kamenshchik, A.Y. and Regoli, D.: J. Cosmol. Astropart. Phys.
\textbf{0802}(2008)015; Aref'eva, I.Y., Bulatov, N.V. and Vernov,
S.Y.: Theor. Math. Phys. \textbf{163}(2010)788.

\bibitem{9} Bamba, K., Nojiri, S. and Odintsov, S.D.: Phys. Rev. D
\textbf{77}(2008)123532; Elizalde, E. and L´opez-Revelles, A.J.:
Phys. Rev. D \textbf{82}(2010)063504.

\bibitem{10} Myrzakulov, R., Saez-Gomez, D. and Tureanu,
A.: Gen. Relativ. Gravit. \textbf{43}(2011)1671; de la Cruz-Dombriz,
A. and Saez-Gomez, D.: Class. Quantum Grav. \textbf{29}(2012)245014.

\bibitem{11} Jamil, M., Momeni, D., Raza, M. and Myrzakulov, R.: Eur. Phys. J. C \textbf{72}(2012)1999;
Bamba, K., Myrzakulov, R., Nojiri, S. and Odintsov, S.D.: Phys. Rev.
D \textbf{85}(2012)104036.

\bibitem{12} Deffayet, C. and Woodard, R.P.: J. Cosmol. Astropart. Phys. \textbf{0908}(2009)023;
Elizalde, E., Pozdeeva, E.O. and Vernov, S.Y.: Phys. Rev. D
\textbf{85}(2012)044002; Class. Quantum Grav.
\textbf{30}(2013)035002; Elizalde, E., Pozdeeva, E.O., Vernov, S.Y.
and Zhang, Y.:  J. Cosmol. Astropart. Phys. \textbf{07}(2013)034.

\bibitem{13} Kamenshchik, A.Y., Tronconi, A. and Venturi, G.: Phys. Lett. B \textbf{702}(2011)191.

\bibitem{14} Kamenshchik, A.Y., Tronconi, A., Venturi, G. and Vernov, S.Y.: Phys. Rev. D \textbf{87}(2013)063503.

\bibitem{15} Bergmann, P.G.: Int. J. Theor. Phys. \textbf{1}(1968)25;
Cliftona, T., Ferreiraa, P.G., Padillab, A. and Skordis, C.: Phys.
Reports \textbf{513}(2012)1.

\bibitem{15*} Lidsey, J.E. et al.: Rev. Mod. Phys. \textbf{69}(1997)373;
Copeland, E.J., Sami, M. and Tsujikawa, S.: Int. J. Mod. Phys. D
\textbf{15}(2006)1753; Frieman, J.A., Turner, M.S. and Huterer, G.:
Ann. Rev. Astron. Astrophys. \textbf{46}(2008)385.

\bibitem{16} Sharif, M. and Waheed, S.: Int. J. Mod. Phys. D. \textbf{21}(2012)1250055; J. Phys. Soc. Jpn.
\textbf{81}(2012)114901; J. Exp. Theor. Phys. \textbf{115}(2012)599.

\bibitem{17} Collins, C.B., Glass, E.N. and Wilkinson, D.A.: Gen. Relativ. Gravit. \textbf{12}(1980)805.

\bibitem{18} Sharif, M. and Zubair, M.: Astrophys. Space Sci. \textbf{330}(2010)399;
Yadav, A.K., Pradhan, A. and Singh, A.K.: Astrophys. Space Sci.
\textbf{337}(2012)379.

\bibitem{19} Sharif, M. and Waheed, S.: J. Cosmol. Astropart. Phys.
\textbf{02}(2013)043.

\bibitem{20} Kucukakca, Y., Camci, U. and Semiz, I.: Gen. Relativ. Gravit. \textbf{44}(2012)1893.

\bibitem{21} Demianski, M., Piedipalumbo, E., Rubano, C.
and Scudellaro, P.: Astronomy and Astrophys. \textbf{481}(2008)279;
Andrianov, A.A., Cannata, F. and Kamenshchik, A.Y.: J. Cosmol.
Astropart. Phys. \textbf{10}(2011)004; Rasouli, S.M.M., Farhoudi, M.
and Sepangi, H.R.: Class. Quantum Grav. \textbf{28}(2011)155004.

\end{thebibliography}
\end{document}